\documentclass[]{elsart}

\usepackage{harvard}

\usepackage{graphicx}

\usepackage{amssymb}

\begin{document}

\begin{frontmatter}



\title{Solar Flare Hard X-ray Spectra Possibly Inconsistent
with the Collisional Thick Target Model}
\author{Eduard P. Kontar and John C. Brown}
\ead{eduard@astro.gla.ac.uk, john@astro.gla.ac.uk}
\address{Department of Physics \& Astronomy, University of Glasgow, G12 8QQ, United Kingdom}



\begin{abstract}
Recent progress in solar Hard X-ray (HXR) observations with
{RHESSI} data and  methods for spectral inversion allow us to
study model-independent mean electron flux spectra in solar
flares. We report several hard X-ray events observed by RHESSI in
which the photon spectra $I(\epsilon)$ are such that the inferred
source mean electron spectra are not consistent with the standard
model of collisional transport in solar flares. The observed
photon spectra are so flat locally that the recovered mean
electron flux spectra show a dip around 17-31 keV. While we note
that alternative explanations, unrelated to electron transport,
have not been ruled out, we focus on the physical implications of
this tentative result for the collisional thick-target model.
\end{abstract}

\begin{keyword}
Sun, Solar Flares, Hard X-rays, Energetic Particles
\end{keyword}

\end{frontmatter}

\section{Introduction}

It has long been recognized (Brown, 1971; Brown and Emslie, 1988;
Johns and Lin, 1992) that even spatially integrated spectra
$I(\epsilon)$ of flare hard X-ray (HXR) collisional bremsstrahlung
bursts carry crucial information on flare electron acceleration,
transport, and energy budget. Even the absence of spectral
features, such as in an (energy-scale-free) pure power-law, is
likely to indicate stochastic multi-scale processes. On the other
hand, any features which exist in the acceleration spectrum are
smeared out in the propagation and radiation process so that any
features detected in $I(\epsilon)$  have potentially very strong
implications for the propagation (Kontar et al, 2003) and
acceleration processes, and for the electron energy budget in
terms of any low energy cut-off.

The large data base of high resolution spectra from the RHESSI
mission (Lin et al, 2002) has made possible a thorough search for
such special HXR spectral diagnostic properties. Care has to be
taken, however, not to misinterpret features as entirely real
solar ones, such as the $50$ keV electron spectral feature in the
July 23, 2002 event (Piana et al, 2003). This can be attributed
partly to pulse pile-up at the high count rate in this event. Two
approaches are used in interpreting $I(\epsilon)$ data. One is to
adopt a parametric form for the mean radiation source electron
spectrum (or for the electron 'injection' spectrum ) and predict
$I(\epsilon)$, using an accurate bremsstrahlung cross-section
$Q(\epsilon,E)$, and search for the best fit in the model
parameter space. This is commonly done using a source-mean flux
spectra ${\overline F}(E)$ containing an isothermal Maxwellian,
$F_T(E)$, defined by an emission measure $EM$ and temperature $T$,
plus a non-thermal $F_{pow}(E)$ parameterized as a single or
double power-law with a sharp low energy cut-off, $E_c$ (Holman et
al, 2003). Such best fits tend to yield an $E_c\sim 10-30$ keV,
well above $kT\sim 1-3$ keV, but with $EM$ large enough that $E_c$
is 'buried' in the $F_T(E)$ 'tail' and it is impossible to be sure
whether $F_{pow}(E)$ has an actual cut off or blends smoothly with
$F_{T}(E)$, especially if there is some spread in $T$. Recently,
Schwartz et al (2003) have reported, for the August 20, 2002 flare
that a best fit requires an $E_c\approx 30$ keV in $F_{pow}(E)$
seen clearly above the isothermal best fit $F_T(E)$. To evaluate
whether such a real feature/cut-off is demanded by the data, as
opposed to simply being compatible with it, it is essential to
adopt a non-parametric approach to interpreting $I(\epsilon)$ -
i.e. to infer from $I(\epsilon)$ what range  of functions
$\overline F(E)$ allows a statistically acceptable fit to
$I(\epsilon)$. This inverse/inferential approach has been adopted,
developed and applied to data by a number of authors (Craig and
Brown 1986, Johns and Lin, 1992; Thomson et al., 1992; Piana et
al, 2003; Kontar et al, 2004; Kontar et al, 2005).

In this paper we apply one of the best available inversion
algorithms of Kontar et al (2004) to several events we have found
in the RHESSI database to show real gaps or dips in $\overline
F(E)$ probably demanded by, and not just consistent with, the
data. The proposed implications of these results for flare
electron acceleration, propagation, and energy budget are briefly
discussed.

\section{Bremsstrahlung Source Models and Constraints on the Emission Spectra}

\subsection{Mean source electron spectra $\overline F(E)$}

In this section we generalize to arbitrary cross-section
$Q(\epsilon, E)$ discussions of bremsstrahlung spectrum
constraints for models discussed earlier by various authors.

Following Brown (1971) and Brown, Emslie and Kontar (2003) we
emphasize that the electron distribution function which can be
inferred from $I(\epsilon)$ without source model assumptions
(apart from optical thinness and isotropy) is the density weighted
mean radiation source electron flux spectrum $\overline F(E)$
(electrons cm$^{-2}$ s$^{-1}$ keV$^{-1}$) defined by
\begin{equation}
\bar F(E)= \frac{1}{\bar n V} \int_V F(E,{\bf r}) \, n({\bf r}) \,
dV.
\end{equation}
where $F(E,{\bf r})$, $n({\bf r})$ are the local electron flux
spectrum and source proton density at position ${\bf r}$ in
radiating volume $V$ with mean target proton density $\bar n =
V^{-1}\int n({\bf r}) \, dV$. $\overline F(E)$ uniquely defines
the bremsstrahlung spectrum at the earth by
\begin{equation}
I(\epsilon) = {1 \over 4\pi R^2} \, \, \bar n V
\int_\epsilon^\infty  \bar F(E) \, Q(\epsilon,E)\, dE, \label{def}
\end{equation}
where $Q(\epsilon, E)$ is the cross-section differential in
$\epsilon$ (Haug, 1997). In general $Q(\epsilon, E)$ should
include angle averaging to allow for anisotropy of the electron
distribution function. The only correction we do not discuss
explicitly here is for photospheric back scatter, though it might
be important for our results.

The general approach to Equation (\ref{def}) is to treat it as an
integral equation to reconstruct $\overline F(E)$ from
$I(\epsilon)$ by deconvolution  through $Q(\epsilon, E)$ using
regularization techniques e.g. (Craig and Brown, 1986; Kontar et
al, 2004). Limitations can be placed on $I(\epsilon)$ for a
physically acceptable solution of (\ref{def}) to exist. In
particular, generalizing (Brown and Emslie, 1988),
\begin{eqnarray}
\frac{d[\epsilon I(\epsilon)]}{d\epsilon} = - {\bar n
V}\left.{\bar F(E) \epsilon Q(\epsilon,E)\,\over 4\pi R^2}
\right|_{E=\epsilon}+\cr\;\;\;\;\;\;\;\;\;\;\;\; +{{\bar n V}
\over 4\pi R^2} \int_\epsilon ^\infty  \bar F(E) \, \frac{\partial
[\epsilon Q(\epsilon,E)]}{\partial \epsilon}\, \mbox{d}E.
\label{inequality}
\end{eqnarray}
Since any physical $\overline F(E)$ should be nonnegative and
since $\partial(Q(\epsilon,E)\epsilon)/\partial \epsilon<0$
$\forall \epsilon >0$ for the relativistic cross-section of Haug,
(1997), we immediately have that $d(I(\epsilon)\epsilon
)/{d\epsilon}<0$, $\forall \epsilon >0$. It is useful to consider
the shape of $I(\epsilon)$ in terms of the local spectral index
$\gamma (\epsilon)$ defined by
\begin{equation}
\gamma (\epsilon)\equiv-\frac{\epsilon}{I(\epsilon)}\frac
{dI(\epsilon)}{d\epsilon}, \label{gamma}
\end{equation}
(note that definition ($\ref{gamma}$) does not correspond to
$I(\epsilon)\sim \epsilon^{-\gamma (\epsilon)}$ except for constant
$\gamma$ (Conway et al, 2003)). Then in terms of the spectral
index the condition $d(I(\epsilon)\epsilon)/{d\epsilon}<0$ can be
expressed
\begin{equation}
\gamma (\epsilon)>1 ,\;\;\;\; \forall \epsilon  \label{gamma1}
\end{equation}
i.e. any bremsstrahlung emission spectrum $I(\epsilon)$ should decrease with a
logarithmic gradient larger than $1$. If the condition
(\ref{gamma1}) is violated then the spectrum $I(\epsilon)$ is not
from an optically thin bremsstrahlung source.

To date no HXR spectrum from a solar flare violating Condition
(\ref{gamma1}) has ever been reported, so there is no challenge to
the belief that flare HXR burst spectra are consistent with
optically thin bremsstrahlung. However, more stringent spectral
compatibility conditions apply to specific models of how $\bar
F(E)$ is formed from an injection spectrum $F_0(E_0)$.

\subsection{Thick target model}

A {\it thick target} is a source in which electrons injected into
the source decelerate to rest under the action of energy loss
processes, within the observed integration time and volume. The
terminology 'thick' is similar to optical depth, a thick target
being a medium where plasma electron column density is sufficient
to stop energetic electrons, or, equally, the mean free path of
the fast electrons is much less than the system length. In other
words, for energy loss cross section $Q_E= -d \log (E)/dN$,
$NQ_E\gg 1$ within the column density $N$ of $V$ at all $E$ of
interest. The resulting mean electron flux spectrum is then the
density-weighted average of the flux spectrum of the electrons
along their energy loss path $z$, as governed by $Q_E$ for all the
energy loss processes involved.

The assumptions in inferring $F_0(E_0)$ from the total $I(\epsilon
)$ are that : (a) the bremsstrahlung emission from acceleration
region can be neglected, and (b) physically distinct
'acceleration' and 'target' (deceleration) regions exist.

Let us consider the motion for an electron in the target from an
injection spectrum $F_0(E_0)$ ($E(z=0)=E_0$) for a given
background plasma density profile $n(z)$, and introduce the column
depth $N(z)$ of the plasma so that $n(z)=dN/dz$. It is important
to note that $z$ is an electron path. Therefore, a column depth
can be also defined as $dN(z)=n(z)v(z)dt$. Comparing with the
original thick target model (Brown, 1971) where only collisional
losses have been taken into account we get, for that case
\begin{equation}\label{eq1}
  \frac{dE}{d N}_{coll}=-K/E
\end{equation}
where $K=2\pi e^4 \Lambda$, $e$ is the electron charge,and
$\Lambda\approx 20-25$ is the Coulomb logarithm.

Assuming that we can find the solution of the equation of motion
of an electron in the form $E=E(E_0,N)$ we can write down the mean
electron flux in the following form
\begin{equation}
\bar F(E)= \frac{A}{\bar n V} \int_{0}^{\infty} F(E(E_0,N))\ dN.
\label{fbar1}
\end{equation}
where $A$ is the injected area. The evolving electron spectrum
$F(E(E_0,N))$ can then  be expressed in terms of the initial or
injected electron spectrum using the continuity equation
$F(E(E_0,N))dE=F_0(E_0)dE_0$. After the change of
variables $N\rightarrow E_0$ in (\ref{fbar1}) one has for the
mean electron spectrum with arbitrary energy losses.
\begin{equation}
\bar F(E)= -\frac{A}{\bar n V} \int_{E}^{E_0(N=\infty,E)}
F_0(E_0)\left(\frac{dE}{dN}\right)^{-1}dE_0. \label{fbar2}
\end{equation}
 (The
mean electron flux given by (\ref{fbar2}) makes use of
deterministic electron propagation ignoring collective and
diffusion effects). In previous papers (Brown and MacKinnon, 1985)
it was also assumed that the upper limit in the integral
(\ref{fbar2}) is infinity \footnote{Actually
$E_0(N\rightarrow\infty,E)=\infty$ for any non-collisional losses
described by the equation of the type $\frac{dE}{d N}=-\varphi
(E)$}, e.g. $E_0(N\rightarrow\infty,E)=\infty$, which is true for
purely collisional losses. Equations (\ref{eq1}) and (\ref{fbar2})
become
\begin{equation}
\bar F(E)= \frac{AE}{K\bar n V} \int_{E}^{\infty} F_0(E_0)dE_0.
\label{fbar3}
\end{equation}

\subsection{Purely collisional thick target}

We now consider the case of a purely collisional thick target
(\ref{fbar3}), when the injection spectrum takes the form
\begin{eqnarray}
F_0(E_0)= - \frac{K\bar n V}{A}\left.\frac{d}{dE} \frac {\bar
F(E)}{E}\right|_{E=E_0} =\cr=K\frac{\bar n V}{A} \frac{\bar F}{E_0^2}
\left(1+\delta (E)\right)_{E=E_0}\label{f0}
\end{eqnarray}
where $\delta (E)=-{d\ln \bar F}/{d\ln E}$ is the spectral index
of the density weighted mean electron flux distribution.
 From equation (\ref{f0}) it follows that, for a
photon spectrum $I(\epsilon)$ to be produced by a purely
collisional thick target with a physical (nonnegative) injection
spectrum $F_0(E_0)\geq 0$, the mean electron flux $\bar F (E)$
derived from the photon spectrum $I(\epsilon)$ must have a
logarithmic slope
\begin{equation}\label{delta}
\delta (E)\geq -1\;\;\;\;\;\; \forall E\neq 0
\end{equation}
(or $d\log \bar F/d\log E \le +1$).

In the next section we report RHESSI $I(\epsilon)$ data in which
the above condition appears to be violated, and a purely
collisional thick target interpretation is impossible.

Equation (\ref{fbar3}) also explicitly says that if the injected
spectrum $F_0(E_0)$ has a low energy cut-off at energy $E_0$, then
the purely collisional thick target mean spectrum $\bar F (E)$ can
have a spectral index $\delta (E)= -1$, for $E< E_0$ i.e. a
logarithmic slope of $\ge+1 $) which is the lowest $\delta$ value
that can be explained by a collisional thick target. However, if
the spectral index at some point is less than -1, then condition
(\ref{delta}) is violated, and non-collisional models must be
involved.

\section{RHESSI observations}

There are flares observed by RHESSI (Lin et al, 2002) with
unusually flat photon spectra resulting in peculiar mean electron
distribution functions ${\bar F}(E)$. Unlike the July 23, 2002
flare (Piana et al, 2003), in these flares the count rate in the
front segments of the RHESSI detectors were too low for pulse
pile-up (Smith et al, 2002) to be important. Forward fits applied
to this data show that the X-ray spectra can be best fitted by a
broken power-law with a low energy cutoff in the range between 15
and 35 keV. This suggests that there are no injected nonthermal
electrons in the range below a few tens of keV. To verify this we
used a recently developed regularization algorithm (Kontar et al.
2004) to infer model - free electron spectra for these events. The
mean electron flux spectrum inferred shows a clear dip at energies
between 15 and 35 keV. A similar feature, though less clear and at
the energies around $50$keV, has been recovered by Piana et al
(2003). The main results are presented in Fig 1-3.

\subsection{Inversion of photon spectra}
The observed spectrum of the flare is the convolution of the
cross-section and mean electron flux. The problem of inferring the
mean electron spectrum is ill-posed (Craig and Brown, 1986). Thus,
one has to avoid unphysical behaviour in ${\bar F} (E)$ using some
constraints. This process is called regularization.

To infer the mean electron spectrum we used the following recently
developed regularization algorithm of Kontar et al, (2004)

\begin{equation}\label{mproblem}
\mathcal{L}(\bf{\overline{F}})=\|{\bf{A}}{\bf{\overline{F}}}-{\bf{I}}\|^2+
\lambda \|{\bf{L\overline{F}}}\|^2,\;\;\;
\mathcal{L}(\bf{\overline{F}})=\mbox{min}
\end{equation}
where $\lambda$ is a regularization parameter, ${\bf L}$ is a
regularizing first order derivative operator, and ${\bf A}$ is the
matrix representation of the cross-section operator $A$ in
\begin{equation}\label{b1}
(A{\overline{F}})(\epsilon)\equiv\frac{1}{4\pi R^2} \,
{\overline{n}} \, V \int_{\epsilon}^{\infty} {\overline{F}}(E) \,
Q(\epsilon,E) \, dE,
\end{equation}
i.e. comprises binned integral components $\int_{\Delta
E_j}Q(\epsilon_i,E)\mbox{d}E$. For our calculations we used the
Haug (1997) cross-section.

One should note that if we try to infer the electron spectra
without additional constraints, e.g. setting to zero the penalty
term in Eq. \ref{mproblem}, the resulting $\bar F(E)$ would be a
violently oscillating function. Such unphysical results come from
the ill-posedness of the problem, since small noise perturbations
in $I(\epsilon)$ will be strongly amplified.

\subsection{Mean Electron Spectrum and Spectral Index}

For our analysis we have selected three flares with unusually flat
local spectra. The data reduction has been performed in the same
way as in Kontar et al (2005). Then the regularization algorithm
has been applied to find the mean electron flux and its spectral
index. Figures 1-3 show flat mean electron flux spectra with a
positive derivative in the range 15-35 keV. The depth of the dip
and the resulting electron spectral index is different from flare
to flare, though qualitatively similar. The April 25 event (Figure
1) has a dip at 19 keV but the error bars do not allow us to
conclude that this dip is real.
\begin{figure}
\begin{center}
  \includegraphics[width=120mm]{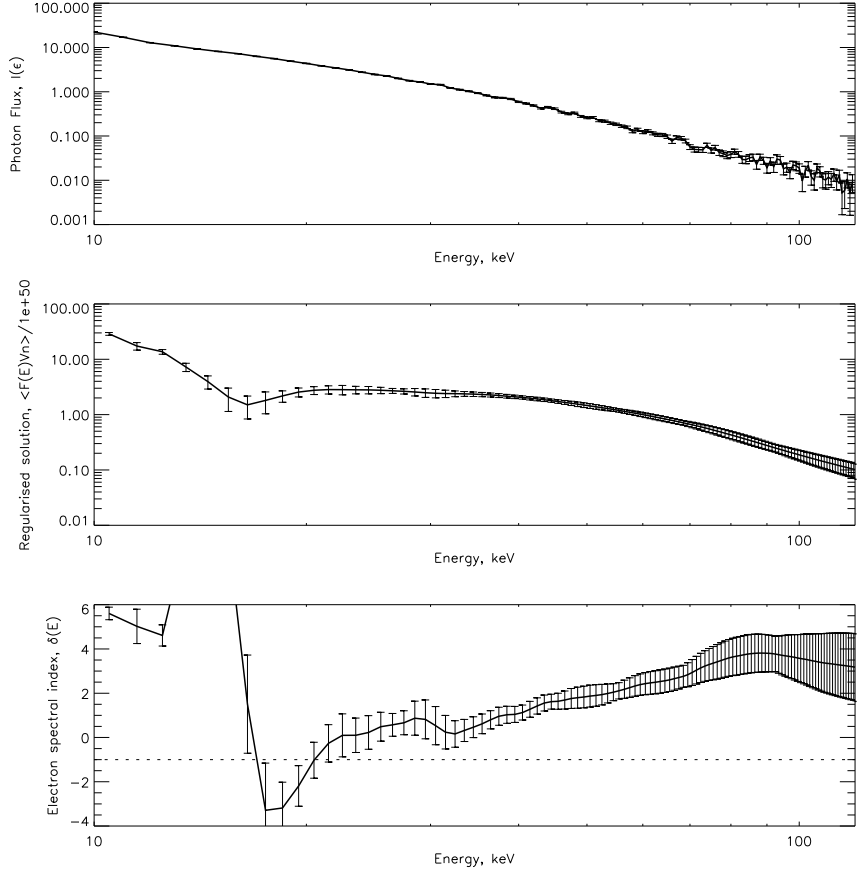}
  \end{center}
\caption{Spatially integrated photon flux $I(\epsilon)$ (upper
panel), mean electron flux ${\overline F}(E)$ and electron
spectral index ${\delta}(E)$ (lower panel) for 25 April 2002
~05:55~UT flare using first order regularization. The photon
spectrum has been accumulated over the impulsive phase of the
flare.} \label{apr25_1}
\end{figure}

The confidence intervals on the regularized solution and spectral
index (Figures 1-3) have been calculated as a maximum deviation
from the solution in a set of 30 random realizations of photon
data within a $\pm 1\sigma$ range. The distribution of errors has
been discussed in Piana et al 2003; Kontar et al, 2004. To allow
for instrumental uncertainties, $1\sigma$ was taken to be not less
than $3\%$ of the photon flux, and thus can be treated as an upper
estimate of the error.

The August 2002 (Figure 2) flare shows an extremely clear gap
around $20$ keV in the mean electron flux. Forward-fit to the data
shows an unambiguous low energy cut-off at $\sim 31$ keV
(Kasparova et al, 2005). To verify the reality of this feature, we
used different orders of regularization for the inversion of the
corresponding flare X-ray spectra. All zero, first, and second
order regularization methods produce the dip (Figure 2).

Figure 3 also shows that at least three successive points have
$\delta$ less than $-1$  with $1\sigma $ confidence and 4 points
less than zero. The former suggests that the thick-target model is
unacceptable while the latter indicates a low energy cut-off in
the injected electron spectrum. Though we have used the $1\sigma$
criterion, the result is much more significant than that $65\%$ of
a single $1\sigma$ deviation since there are three successive such
deviations. While a more rigorous statistical analysis is
necessary, this should be significant at roughly the level of
$1-(1-0.65)^3$ or about $97\%$ which makes it highly suggestive at
least.

\begin{figure}
\begin{center}
  \includegraphics[width=120mm]{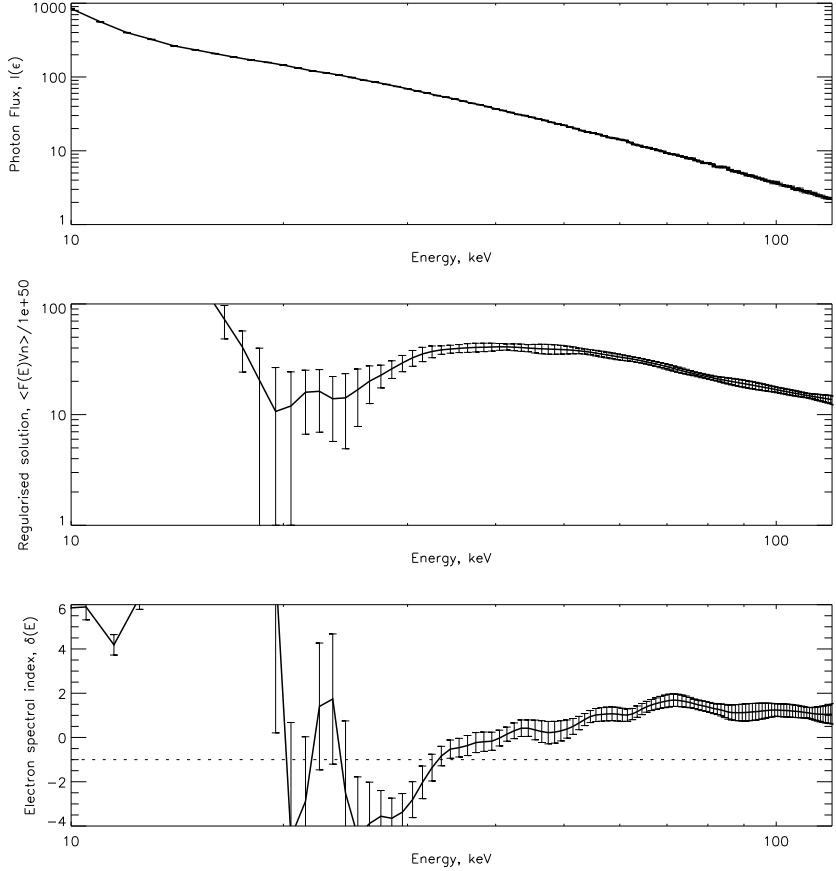}\\
  \end{center}
\caption{The same as in Fig 1 for 20 August 2002 ~08:24~UT flare.}
\label{aug20_1}
\end{figure}

\begin{figure}
\begin{center}
  \includegraphics[width=120mm]{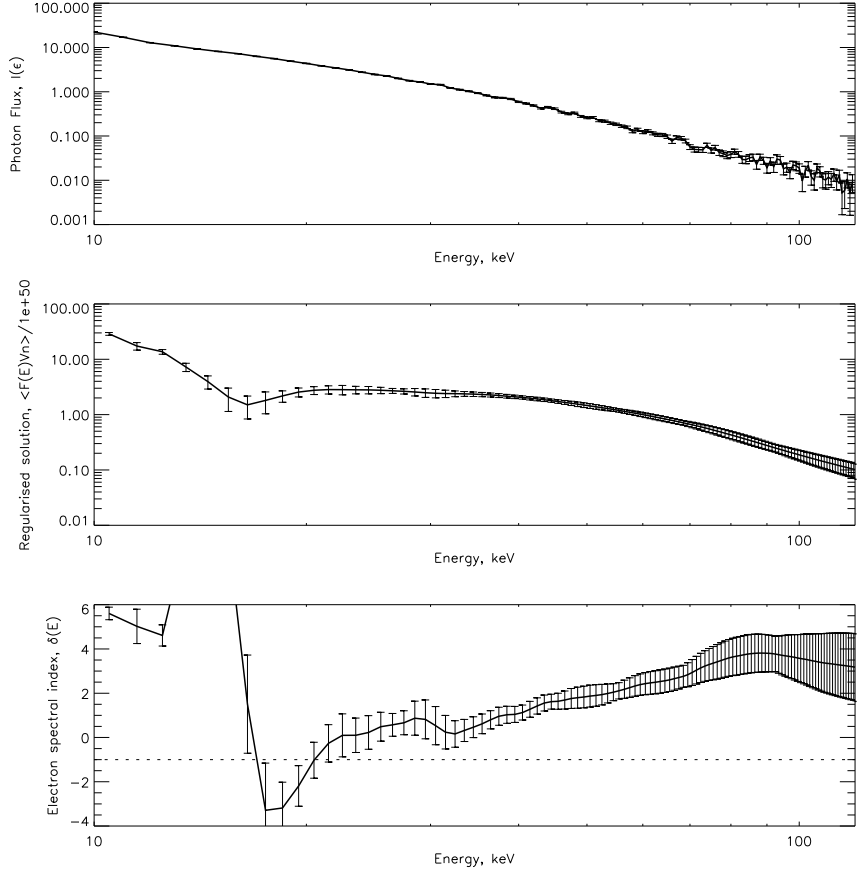}\\
  \end{center}
\caption{The same as in Fig 1 for 17 Sept 2002 ~05:55~UT flare.}
\label{sep17_1}
\end{figure}

\section{Discussion and conclusions}

Our analysis of the observed photon spectra shows features that
have not been observed before. Johns and Lin (1992) reported the
downturn of the derived electron spectrum for earlier balloon
data, though the errors were too big to be conclusive. In the
results presented in this paper we are much more confident that a
dip in the electron distribution is required by the photon
spectra, though we caution that the photon spectrum itself is
subject to additional corrections for photospheric albedo,
directivity, and instrument effects (Smith et al, 2002).

One of the important implications of our result if upheld is
related to the accelerated spectrum $F_0(E_0)$. As can be seen
from the results, the photon spectra require a low energy cut-off
in the injected spectra $F_0(E_0)$ to be produced, e.g. the
absence of electrons in the range below 20-30 keV. This is quite a
strong requirement on acceleration models, that often produce
extended power law spectra above thermal energies of a few keV.
The cut-off poses physical challenges to electron acceleration and
transport since an electron beam with a velocity distribution
function such that $f(v)\sim \bar F(v^2)$ should be unstable to
the generation of plasma waves if $df(v)/dv>0$.

We stress that, if the existence of a low energy cut-off in the
injected electron spectrum $F_0(E_0)$ is confirmed, its
implications for total flare electron power, acceleration and
propagation are profound. Moreover, the evidence of a steep turn
over in ${\bar F}(E)$ would reject the long-standing collisional
transport thick-target model of flare electron propagation (Brown,
1971) where the accelerated electron spectrum $F_0(E_0)$ is
modified solely by Coulomb collisions along the electron path.

{\bf Acknowledgements} We are grateful for the support of a PPARC
Grant and to A.G. Emslie, C. Johns-Krull, J. Kasparova, R.P. Lin,
A.M. Massone, M.Piana, and R.A. Schwartz for numerous stimulating
discussions.

\end{document}